\documentstyle[12pt,psfig]{article}

\def\appendix{\par
 \setcounter{section}{0}
 \setcounter{subsection}{0}
 \def\thesection{Appendix \Alph{section}}}
 
\font\blackboard=msbm10 at 12pt
\font\blackboards=msbm7
\font\blackboardss=msbm5
\newfam\black
\textfont\black=\blackboard
\scriptfont\black=\blackboards
\scriptscriptfont\black=\blackboardss

\newcommand{\gone}[1]{}
\begin{document}
\pagestyle{plain}
\setcounter{page}{1}
 
\baselineskip16pt
 
\begin{titlepage}
 
\begin{flushright}
NSF-ITP-97-141\\
hep-th/9711097\\
\end{flushright}
\vspace{20 mm}

\begin{center}

{\Large {\bf  The Shape of Branes Pulled by Strings
}}

\bigskip

\centerline{Akikazu Hashimoto}

\bigskip

{\it Institute for Theoretical Physics \\
University of California\\
Santa Barbara, CA 93106-4030\\}

\end{center}

\bigskip

\begin{abstract}
\noindent We examine the system where a string stretches between pair
of D-branes, and study the bending of the D-brane caused by the
tension of the string.  If the distance between the pair of D-branes
is sent to infinity, the tension of the string stretching between them
is strong enough to pull the spike all the way to infinity.  We study
the shape of these spikes when the branes are finite distance apart
using two different methods. First, we consider a string stretched
between a pair of D2-branes in type IIA theory by going to the
M-theory limit in which all of these branes are M-theory 2-branes
embedded along a holomorphic curve.  Second, we consider a D-string
stretched between a pair of D3-branes in type IIB theory and infer the
geometry of the D3-brane embeddings from the configuration of the
adjoint scalar field in the magnetic monopole solution of Prasad and
Sommerfield.  The case of fundamental string stretching between a pair
of D3-branes follows from S-duality.  The energy of these
configurations matches the expected value based on fundamental string
and D-string tensions.
\end{abstract}

\begin{flushleft}
November 1997
\end{flushleft}
\end{titlepage}
\newpage

\section{Introduction}

In the recent years, brane objects have emerged in the forefront as
the key ingredient in the study of string theory and field theories.
Branes are extended objects in space-time; when they occupy
$p$-dimensional volume in space, they are referred to as $p$-branes.
Fundamental particles and strings are also branes, in a sense that
they could be thought of as 0-branes and 1-branes. Other types of
branes can appear as a soliton of the theory, and can be classified
according to its dimension and its charge. Some of these branes
admits a simple perturbative description as D-branes
\cite{polchinski}.

These brane objects have found spectacular range of applications.  One
active avenue of investigation in the recent years have been the brane
construction of quantum field theories.  The basic idea stems from the
fact that the world volume of D-branes contains a U(1) gauge field in
its massless open string spectrum. When $N$ parallel D-branes approach
each other, the $U(1)^N$ gauge symmetry is enhanced to $U(N)$ since
the ``W-bosons'' corresponding to strings stretching between the
branes also become massless \cite{ed}. Various matter content can be
added to the theory, for example by including orthogonally
intersecting branes \cite{bsv1,hw96,egk97}.

A configuration of interest from the point of view of the brane
construction of gauge theories is the one where an open $p$-brane ends
on another brane \cite{as}.  One can describe, for example, a magnetic
monopole in 3+1 dimensional $SU(2)$ gauge theory in terms of a pair of
parallel D3-brane with an orthogonal D1-brane ending on the D3-branes
\cite{GrnGutp96,Diaconescu}.  Other configuration of branes stretching
between branes can also be dynamically generated when certain class of
orthogonal D-branes cross as they move past one another in space
\cite{hw96,egk97,BDG97,DFK97,BGL97}.

The branes which appear in string theory generally have tensions of
order $1/g$ or $1/g^2$ where $g$ is the string coupling constant.
Therefore, for arbitrarily small but finite $g$, these branes have
finite tension and will bend slightly when some other brane is
attached and exerts its tension.  This effect was studied
systematically for the system consisting of a pair of Neveu-Schwarz
five-branes in type IIA theory, suspending a collection of open
D4-branes with endpoints along the Neveu-Schwartz five brane world
volume \cite{witten4d}.  The bending of the NS5-brane far away from
the D4-brane was given a natural interpretation as running of the
coupling constant for the D4-brane world volume theory.  The geometry
of NS5-brane near the D4-brane was studied by taking advantage of
M-theory limit in which both the NS5-brane and the D4-brane becomes an
M-theory five brane.  BPS configurations in this limit corresponds to
holomorphic embeddings of the world volume into space time
\cite{BBS95}. Such an embedding is generically smooth, but by
shrinking the radius of the 11-th dimension, this smooth configuration
degenerates into a singular geometry of D4-branes with endpoints on
Neveu-Schwarz 5-branes. (A closely related configuration of D2-brane
stretching between NS5-branes and D4-branes was discussed in
\cite{brodie97}.)

In this article, we attempt to address the question ``what is the
shape of D-branes which are pulled by strings.'' In other words, we
will consider a system where an open string has one of its endpoints
on a D-brane\footnote{The same system in the perturbative context was
also studied in \cite{Giddings96}. Similar situation arises also in
the context of topological defect solutions in ordinary field theories
\cite{CarrollTrodden}.}, and study the bending of the D-brane in
response to the tension exerted by the string.  We will work in the
context of weakly coupled string theory, where the brane geometry is
expected to contain singularities.  Understanding the nature of this
singularity will be one of the goals of this paper.  To be concrete,
we will focus on the geometry of the bending of D3-branes due the
tension of either a fundamental string or a D-string which ends on the
worldvolume of the D3-brane \cite{GrnGutp96,Diaconescu}.

An important step in addressing this question was made in recent
papers \cite{CM97,Gibbons97}. (See \cite{hlw97,LPT97,Thorlacius97} for
related work.) These authors considered an electric BPS field
configurations of the Dirac-Born-Infeld (DBI) action ($p=3$ for
D3-brane)
$$ S =  T_p \int d^{p+1} x\  \sqrt{-\det(\eta_{\mu \nu} + T^{-1} F_{\mu \nu} + \partial_\mu X^i \partial_\nu X_i)}.$$
These electric charges correspond to endpoints of fundamental
strings. The BPS configuration they found was of the form (in Coulomb
gauge)
$$T^{-1} A_0 = c_p/r^{(p-2)}, \qquad X^9 = c_p/r^{(p-2)}.$$
Here, $p=3$, $T = (2 \pi \alpha')^{-1}$, and
$$T_p  =  {1 \over g} (2\pi)^{(1-p)/2} T^{(p+1)/2}, \qquad c_p = {g \over (p-2) \Omega_{p-1}} (2\pi)^{(p-1)/2} T^{-(p-1)/2}.$$
The scalar field $X_9$ encodes the embedding of the D3-brane into
space-time which contains an infinite spike at $r=0$ extending in the
$x_9$ direction.  The energy density for this configuration is
$$T_{00}(r) = {c_p^2 (p-2)^2 T_p \over r^{2(p-1)}}.$$
The total energy of this configuration is the integral of this expression over
the 3-brane world volume
$$E = \int d^{p}r {c_p^2 (p-2)^2 T_p \over r^{2(p-1)}} = 
 c_p (p-2) \Omega_{p-1} T_p \int dr {(p-2) c_p  \over r^{p-1}}
.$$
Although this integral is formally diverging near $r=0$, its physical 
origin can be elucidated by making a change of variables
$$d X_9 = dr {(p-2) c_p  \over r^{p-1}}$$
Also noting that $c_p (p-2) \Omega_{p-1} T_p = T$, the expression for the
energy of the system becomes
$$E = T \int dX_9.$$
This expression has natural interpretation as the energy of string
with tension $T$ stretching in the $X_9$ direction.  This lead the
authors of \cite{CM97,Gibbons97} to interpret the semi-infinite spike
on the 3-brane world volume as the string itself.

\begin{figure}
\centerline{\psfig{file=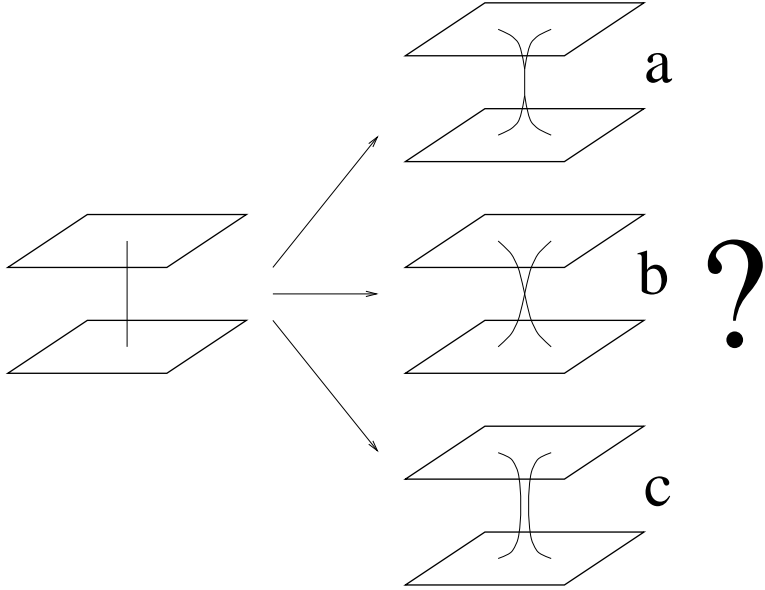,width=4in}}
\caption{A string stretches between pair of parallel D3-branes,
causing the D3-branes to bend.\label{fig1}}
\end{figure}

This picture runs into problems, however, if one wishes to consider a
configuration where a string stretches between a pair of parallel
3-branes separated at a finite distance, say at $X_9=-b$ and $X_9=b$,
respectively.  One might imagine that near $X_9=-b$ or $X_9 = b$, the
brane configuration is given by the BPS configuration described above,
i.e.\ $X_9 = -b+c_3/r$ and $X_9'=b-c_3/r$. The difficulty arisis at
the point in $r$ where $X$ and $X'$ meets, since these curves do not
interpolate naturally into each other. Natural expectation is that in
the presense of the second D3-brane, the geometry of the bending of
the D3-branes is modified near $r=0$.  There are three qualitative
possibility for how this might come about. One possiblity is for the
D3-brane to be pinched off and collapse to zero radius at finite
distance before it meets the other D3-brane (figure \ref{fig1}.a).
One would expect a string to stretch between the points where the
D3-branes are pinched off.  Another possiblity is for the two
D3-branes to meet precicely at the point where the radius shrinks to
zero (figure \ref{fig1}.b). Yet another possiblity is for the pair of
D3-branes to smoothly interpolate each other in a wormhole type
geometry (figure \ref{fig1}.c).

Possibility \ref{fig1}.c has been discussed by authors of
\cite{CM97,Gibbons97}. They considered a cantenoidal static
configuration of the Born-Infeld action of the form
\begin{eqnarray*}
E_r & = & {A \over \sqrt{r^{2p-2}-r_0^{2p-2}}} \\
\partial_r X_9 & = & {B \over \sqrt{r^{2p-2}-r_0^{2p-2}}} \\
\end{eqnarray*}
where the constant $r_0^{2p-2} = B^2-A^2$ controls the size of the
throat, and $A = (p-2) c_p$ is fixed by the charge of the string
endpoint.  Unfortunately, such a configuration can only describe a
3-brane anti-3-brane configuration, as the orientation of the brane
reverses as one interpolates down the throat.

In the remainder of this paper, we will discuss different approaches
for investigating the geometry of the brane configurations illustrated
in figure \ref{fig1}.  Note that since the configuration in figure
\ref{fig1} contains a set of parallel D3-branes, the full non-abelian
generalization of the Born-Infeld action is required in order to study
this system along the lines of \cite{CM97,Gibbons97}. Unfortunately,
no satisfactory formulation of such an action is currently
available. (Tseytlin has recently proposed a concrete formulation of
the non-abelian Born-Infeld action using the symmetrized trace
\cite{NDBI}. Although this proposal passes many tests, some issues
remain in understanding its fluctuation spectrum around a non-trivial
background \cite{akiwati}.)  We will therefore approach this problem
using indirect methods.  We begin by considering more general class of
static solutions for Born-Infeld action which appears to describe
spikes being pinched off to a point and degenerating to a string as
illustrated in figure \ref{fig1}.a. This approach turns out to be
problematic from the point of view of energy. Two alternative methods
for studying the shape of the bending is discussed.  First, we
consider a closely related system of a string stretching between
D2-branes in type IIA theory, where one can exploit known facts from
M-theory\footnote{I am grateful to Juan Maldacena for pointing out
this approach to me.} by extrapolating the coupling constant to strong
coupling. Unfortunately, it is not clear how this approach can be
extended to the case of D3-branes.  The second approach is to consider
the Yang-Mills approximation to the Born-Infeld action.  In this
approach, the shape of the branes is read off from the field
configuration of the Higgs scalar in the Prasad-Sommerfield magnetic
monopole solution \cite{PS75} for the SU(2) gauge theory.  Energies
for these configurations can be computed and gives the expected
results.  In the abelian case, the BPS field configuration for the
Maxwell theory turned out to also satisfy the BPS condition in the
Born-Infeld theory \cite{CM97,Gibbons97}. Perhaps for BPS
configurations, the Yang-Mills and Born-Infeld action is supposed to
give same answers even for non-abelian gauge theories. If true, this
idea may provide a useful guiding principle for the formulation of
non-abelian Born-Infeld actions. (We will show in \ref{app1} that this
property holds for Tseytlin's action \cite{NDBI}.) The fact that the
energies of these configurations work out exactly provides some
support for this idea.

\section{Static solutions of Dirac-Born-Infeld action}

\begin{figure}
\centerline{\psfig{file=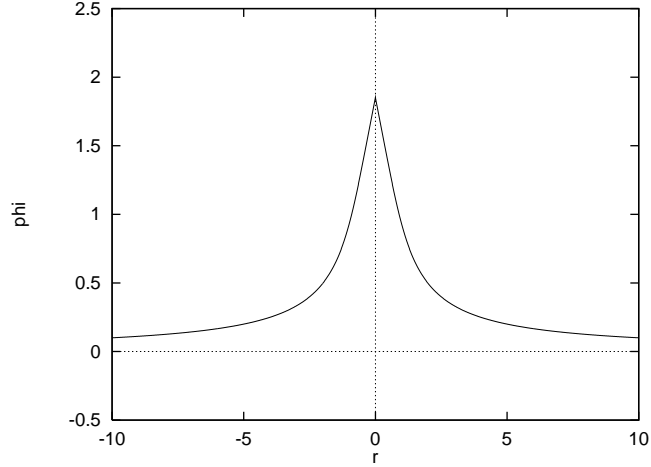,width=3.5in}}
\caption{Electrostatic potential (in units of
$[(p-2)c_p]^{1 \over p-1}$) for a BIon solution in 3+1 dimensional
Born-Infeld theory.\label{fig2}}
\end{figure}

Let us begin by looking for a static configuration of the
Dirac-Born-Infeld action which might capture the geometry of the brane
configuration of figure \ref{fig1}.  For concreteness, let us
concentrate on 3+1 dimensional Born-Infeld theory with 6 scalars.  For
our purpose here, we can set all fields other than the gauge potential
$\phi=T^{-1}A_0$ and one of the transverse scalar $y=X_9$ (say) to
zero. Since we are interested in static configurations, we can also
set all time-derivatives to zero. The statics problem then reduces to
extremization of \cite{Gibbons97}
$$T_p \int d^px \sqrt{ 1 + |\nabla y|^2 - |\nabla \phi|^2 + (\nabla y \cdot \nabla \phi)^2 - |\nabla \phi|^2 |\nabla y|^2}$$
which gives rise to equations of motion
\begin{eqnarray}
\nabla \cdot { -\nabla \phi +\nabla y (\nabla y \cdot \nabla \phi) - \nabla \phi
(\nabla y)^2 \over
\sqrt{ 1-(\nabla \phi)^2 + (\nabla y)^2 + (\nabla y \cdot \nabla \phi)^2
- (\nabla y)^2 ( \nabla \phi)^2 }}&=&0  \label{eqmo1}\\
\nabla \cdot  { \nabla y +\nabla \phi (\nabla y \cdot \nabla \phi) - \nabla y
(\nabla \phi)^2 \over
\sqrt{ 1-(\nabla \phi)^2 + (\nabla y)^2 + (\nabla y \cdot \nabla \phi)^2
- (\nabla y)^2 ( \nabla \phi)^2 }}& =&0 \label{eqmo2}. 
\end{eqnarray}
The approach of this section is motivated by the existence of static
charged field configuration \cite{BI34} (referred to as a ``BIon'' in
\cite{Gibbons97}) with contribution only from the gauge sector:
$$ \phi(r) = \int_r^\infty {(p-2) c_p \over \sqrt{(p-2)^2 c_p^2 + r^{2(p-1)}}},\qquad\qquad y(r)=0.$$
This integral converges everywhere and has the form illustrated in
figure \ref{fig2} (for $p=3$).
If it had been the scalar on the world volume which had the form
illustrated in figure \ref{fig2}, such a solution appears to describe
a configuration similar to the one illustrated in figure
\ref{fig1}.a. It turns out that such a solution can be constructed by
boosting the field configuration in the $\phi$-$y$ space along the
lines explained in \cite{Gibbons97}. With some algebra, it is possible
to summarize the set of solutions discussed in \cite{Gibbons97} as
family of solutions parameterized by single variable $a$:
\begin{equation}
\phi(r) =  \int_r^\infty dr\, {1 \over \sqrt{1 -a + ({r^{p-1}  \over (p-2) c_p})^2}}, \qquad 
y(r) =  \int_r^\infty dr\, \sqrt{a \over 1 -a + ({r^{p-1}  \over (p-2) c_p})^2}.\label{family}
\end{equation}
It is straightforward to verify that these solutions satisfy the
equations of motion (\ref{eqmo1}) and (\ref{eqmo2}). These solutions
have also been arranged so that the electric charge which couples to
$\phi$ at the origin is precisely $c_p$. For the range of parameters
$0<a<1$, the scalar field $y(r)$ essentially takes the form of figure
\ref{fig2}.  At $r=0$, the brane has a cusp, but takes on a finite
value there.  This provides a natural point for a string to end. Two
sets of 3-branes in such a configuration admits a natural continuation
with strings stretching between them, giving rise to a concrete
realization of a configuration illustrated in figure \ref{fig1}.a.
As $a$ approaches 1, $y(r=0)$ grows indefinitely until it reaches the
point $a=1$ at which we recover the BPS configuration $y =
c_p/r^{p-1}$. For $a >1$, these solutions corresponds to the
cantenoids (solutions of type illustrated in figure \ref{fig1}.c
\cite{CM97,Gibbons97}). In this paper, we will focus on the region
$0\le a \le 1$.

The configuration of branes for range of parameters $0 \le a \le 1$
appears to naturally capture the geometry illustrated in figure
\ref{fig1}.a. Even before doing any further analysis, however, we know
that this can't be the proper description of the bending of the
D-brane due to the tension of strings. This is because the lowest
energy configuration of strings stretching between the branes is a BPS
configuration, whereas the solution (\ref{family}) is BPS only for
$a=1$. It is nonetheless quite amusing and instructive to examine the
static energies for these configurations and to understand precisely
how this solution fails to describe the bending of D3-branes.

\begin{figure}
\centerline{\psfig{file=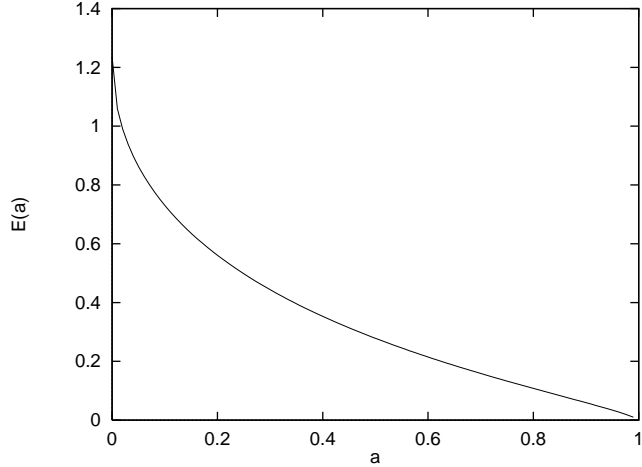,width=3.5in}}
\caption{Combined energy in units of $T \sqrt{c_3}$ of bent D3-brane
and stretched strings for one parameter family of configuration
(\ref{family}) as a function of the parameter $a$. \label{fig3}}
\end{figure}

First, let us take a case where there is only one D3-brane and
consider a string which ends on this D3-brane on one end and the other
end extending infinitely.  If the D3-brane takes on a shape described
given by the solution (\ref{family}), the total energy of the system
will consist of energy density stored on the world volume of the
D3-brane and the energy stored in the string which stretches to
infinity
$$E_{tot} = \int  d^{p}r\, T_{00}(r) + T\int_{y(0)}^\infty dy$$
This expression diverges due to the contribution from the string extending
indefinitely.  A convenient way to subtract this infinite additive constant in energy is to subtract the energy of semi-infinitely extended string
$$E = E_{tot} - T\int_0^\infty dy = \int d^{p}r\, T_{00}(r) - T
\int_0^{y(0)} dy= \int_0^\infty dr\ \left[r^{p-1} \Omega_{p-1}
T_{00}(r)\right] - T y(0)$$
The energy density of this system is given by
$$
T_{00}=T_p \left({1+ (\nabla y)^2 \over \sqrt { 1+|\nabla y|^2 -|\nabla \phi |^2
+ (\nabla y\cdot \nabla \phi) ^2 - (\nabla y)^2(\nabla \phi)^2 }}-1\right).
$$
For the configuration (\ref{family}), the contribution for the energy
density from the world volume of the brane is given by\\
\parbox{\hsize}{
\begin{eqnarray*}
\lefteqn{r^{p-1} \Omega_{p-1}  T_{00}(r)}\\
& = &
\Omega_{p-1}  T_p  (p-2) c_p \left( -\left({r^{p-1} \over (p-2) c_p}\right) + {\frac{a}
    {{\sqrt{1 - a + {\left({r^{p-1} \over (p-2) c_p}\right)^2}}}}}
   + {\sqrt{1 - a + {\left({r^{p-1} \over (p-2) c_p}\right)^2}}}\right)\\
&=&
T \left( -\left({r^{p-1} \over (p-2) c_p}\right) + {\frac{a}
    {{\sqrt{1 - a + {\left({r^{p-1} \over (p-2) c_p}\right)^2}}}}}
   + {\sqrt{1 - a + {\left({r^{p-1} \over (p-2) c_p}\right)^2}}}\right)
\end{eqnarray*}}
where we have used the fact that $\Omega_{p-1} T_p (p-2) c_p =T$. This
expression can be integrated explicitly. Including the $T y(0)$
contribution and setting $p=3$, we find
$$
E(a) = (T \sqrt{  c_3 }){\Gamma({1\over 4})^2 \over 12 \sqrt{\pi}} (\sqrt{a}-2)(\sqrt{a}-1) (1-a)^{-1/4}.$$
The form of this $E(a)$ is illustrated in figure \ref{fig3}, from
which one can clearly observe that the energy is minimized at $a=1$.
This is the point where the spike have been pulled all the way out to
infinity and the brane configuration collapses to the simple
expression $y=c_p / r^{p-2}$. This is to be expected since $a=1$ is
the only BPS solution among the family of configurations parameterized
by $a$.

The physical picture behind this phenomenon is simply that when a
string attaches itself to a D-brane, the pull of the string tension
will create a spike on the world volume of the D-brane. However, for
finite $g$, the tension of the string is strong enough to pull the
spike out of the D-brane indefinitely. If the original strings extends
infinitely in the direction transverse to the brane, then the spike
will grow infinitely in the direction of the string until it assumes
the minimum energy configuration which is simply $y = c_p/r^{p-2}$.
However, since $c_p$ is linear in $g$, the spike becomes narrower in
the weak coupling limit.

This picture leads to a slightly different interpretation of the spike
than the authors of \cite{CM97,Gibbons97}, where because of the
matching of the tension, the spike was interpreted as the string
itself. Here, the spike is to be interpreted as a part of the
D3-brane.  That the tension matches exactly with the tension of
fundamental strings is simply the reflection of the fact that D3-brane
must bend in such a way to generate precisely the tension exerted by
the attached string in order to achieve static equilibrium. However,
the distinction may not be so sharp when the radius of the spike
becomes small.  At length scale much smaller than the radius of the
spike, one can physically identify the spike as being made of D3-brane
by measuring its local Ramond-Ramond charge. At length scales much
larger than the radius of the spike, however, the net Ramond-Ramond
charge is cancelled by the brane on the other side of the tube. On the
other hand, the tube will appear to carry Neveu-Schwarz charge through
coupling with the world volume gauge field background.  So from the
consideration of charges, the spike, at length scales much larger than
its radius, looks more like a string.  Perhaps there is no sense in
which the D3-brane spike can be distinguished from a fundamental
string when the radius of the spike is smaller than the string scale.

With the information at hand, one can qualitatively address the
situation where a string is stretching between pair of D3-branes
separated by finite distance.  The tension of the string will create a
spike on D3-branes at both ends of the string. The tension of the
string is strong enough to pull the D3-branes together until the
length of the string shrinks to zero. Thus, one expects a geometry
like the one illustrated in figure \ref{fig1}.b to be the final
configuration.  One might suppose that this configuration corresponds
to pair of configuration of the form (\ref{family}) for a particular
value of $0 <a <1$ so that the height of the spike is precisely half
the distance between the D3-branes.  The energy of this configuration
can be computed by combining the integrated stress energy tensor from
both the branes.  Energies $E(a)$ illustrated in figure \ref{fig3}
can be interpreted here as the amount of excess energy with respect to
the BPS energy of the brane configuration at the particular value of
$a$.  The fact that $E(a)$ is strictly positive for $0<a<1$ implies
that this is not quite the equilibrium configuration for the
brane. After the spike meets, the brane must adjust its shape slightly
in order to achieve the lowest energy configuration.  In the following
sections, we will attempt to calculate the precise shape to which the
branes relax in achieving such a state.

\section{A view from M-theory}

The main problem with the approach of the previous section was the
fact that we were working with abelian Born-Infeld action although the
system is best described by a non-abelian theory. In this section, we
will consider a closely related problem of understanding the geometry
where a string stretches between a pair of type IIA D2-branes.  For
such a system, one can bypass the need to use the non-abelian
Born-Infeld action by going to the M-theory limit of the theory, where
both the string and the D2-brane becomes an M-theory membrane.  BPS
configuration for these membranes can be realized explicitly as
holomorphic embeddings of the world volume of these branes into
spacetime \cite{BBS95}.

Consider a configuration in type IIA theory where a string stretches
between a pair of D2-brane which extends infinitely in $X_1$ and $X_2$
direction and is separated by some finite distance in $X_3$ direction.
In the M-theory limit, the string becomes a ``tube'' of membrane which
wraps around $X_{10}$ direction and extends in $X_3$ direction. The
entire system of D2-branes and string can be thought of as a single
membrane object.  In order to describe a BPS configuration of this membrane, 
it is convenient to introduce complex coordinates
$$z = X_1 + i X_2, \qquad w = X_3 + i X_{10}.$$
Configuration of the type
$$f(z,w)=0$$
where $f(z,w)$ is a holomorphic function will correspond to a BPS
configuration of a membrane in M-theory.

Since $X_{10}$ is a periodic variable, it is convenient to work 
instead with 
$$t = e^{w/R}$$
where $R$ is the radius of the 11-th direction. Our goal then is to
find a holomorphic function $f(z,t)$ which vanishes along the locus of
the membrane world volume which arise from lifting the configuration
of a string stretched between two D2-branes to M-theory. Finding
$f(z,t)$ satisfying this requirement is analogous to the procedure for
finding a five brane configuration which was described in
\cite{witten4d}, and we will use the same methods here.  In
particular, since there are two branches of membranes extending
infinitely in the complex $z$-plane, we expect $f(z,t)$ to have two
zeros for fixed value of $z$.  Similarly, the membrane wraps the 11-th
direction precisely once. Therefore, for fixed value of $t$, we expect
$f(z,t)$ to have exactly one zero.  Therefore, $f(z,t)$ must be at
most quadratic in $t$ and linear in $z$.  The most general holomorphic
curve satisfying this requirement is
$$f(z,t) = (az+b) t^2 + (cz+d) t + (ez+f).$$
At the zero of $ez+f$, $f(z,t)$ has a zero at $t=0$. In terms of
$w=X_3+iX_{10}$ coordinate, this corresponds to $X_3 = -\infty$ and
implies the presence of a string stretching indefinitely in the $X_3
\rightarrow -\infty$ direction. Similarly, at the zero of $az+b$, the
zero of $f(z,t)$ runs off to infinity. This implies the presence of a
string stretching indefinitely to the $X_3 = +\infty$ direction. Since
we want the strings to stretch only between the two ``would be''
D2-branes, both of these features are undesirable.  They can be
eliminated by setting $a$ and $e$ to zero.  Also, since the overall
scale of $f(z,t)$ is irrelevant, we can set one of the remaining
coefficients, say $f$ to 1.  The $b$ can be absorbed into the scale of
of $t$. Similarly, coefficients $c$ and $d$ can be absorbed into the
choice of scale and the origin of $z$.  The holomorphic embedding of a
membrane in $(z,t)$ plane which reduces to a string stretching between
D2-branes in type IIA theory is therefore given by a curve
$$ t^2 + 2 z t + 1 = 0,$$
or equivalently,
$$z = \frac{1}{2} \left(t + {1 \over t} \right).$$
In terms of the original coordinates, this becomes
\begin{equation}
\begin{array}{rcl}
X_1 &= &\cosh(X_3/R) \cos(X_{10}/R)\\
X_2 &= &\sinh(X_3/R) \sin(X_{10}/R)
\end{array} \label{curve}
\end{equation}
This curve does not preserve rotational invariance in the $(X_1,X_2)$
plane.  This might have been expected since we also broke the
rotational invariance of $(X_3,X_{10})$ plane by compactifying along
the $X_{10}$ direction, and the holomorphic curve relates the two
planes.  In the weak coupling limit where we expect to recover a
rotationally invariant string configuration, this asymmetry also
disappears.

\begin{figure}
\centerline{\psfig{file=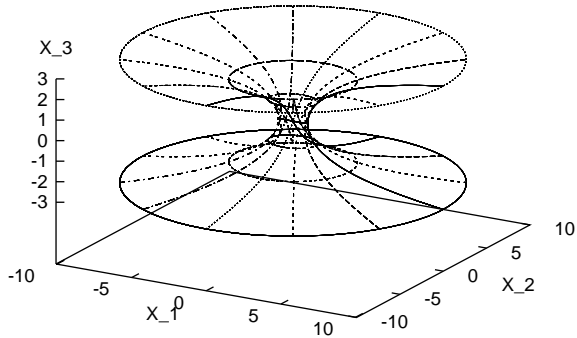,width=3.5in}}
\caption{Projection to $(X_1,X_2,X_3)$ space for the holomorphic curve
for the M-theory description of a string stretching between  a pair of
D2-branes\label{fig4}}
\end{figure}

We illustrate the projection of this curve in the $(X_1,X_2,X_3)$
space in  figure \ref{fig4}.
It has a feature that as $r = |X_1 + i X_2|$ is taken to infinity, the
value of $X_3$ grows like $\ln(r)$. This means that there is no clear
notion of the ``position of the branes at infinity,'' which will
require additional care when taking the limit back to the weakly
coupled type IIA description.

It is also instructive to examine the cross section of this curve with
respect to $X_1=0$ plane or $X_2=0$ plane.  For $X_2=0$, this curve
reduces to
$$ X_1 = \pm \cosh(X_3/R)$$
and appears to describe a geometry similar to that of a cantenoidal
configuration illustrated in figure \ref{fig1}.c. On the other hand,
the cross section with $X_1=0$ plane takes the form
$$X_2 = \pm \sinh(X_3/R).$$
Such a curve appears to cross each other at $X_3=0$.  In the full
M-theory description, however, these branches do not really cross,
since they correspond to different points in the $X_{10}$ direction.
In the limit where weakly coupled type IIA theory is recovered,
however, the $X_{10}$ direction is shrinking and we expect crossing at
$X_3=0$ to degenerate to a pinched brane configuration, similar to the
ones discussed in the previous section. It should be also noted that
when this geometry is projected onto $(X_1,X_2,X_3)$ space, this
apparent crossing is required to undo the orientation reversal of the
branes that would otherwise lead to a brane anti-brane configuration
(figure \ref{fig1}.c).

Let us now imagine taking the limit of shrinking the radius of the
11-th dimension to zero size where we recover the weakly coupled type
IIA description.  The naive $R \rightarrow 0$ limit of (\ref{curve})
appears to collapse the locus of the membrane to $X_3=0$ plane. This
problem is related to the fact that due to the logarithmic divergence
of the brane coordinates, there is no well defined notion of the
distance between the two branches of the curve at asymptotic region $r
\gg 1$. In M-theory, choosing the distance between the brane is
analogous to the choice of scale in asymptotically free theories, and
one must perform a certain renormalization procedure in order to
recover the weakly coupled type IIA description of D2-branes separated
by a finite distance.  One way to achieve this is to require that the
brane passes through the point $(X_1,X_2,X_3) = (\bar{r},0,b)$ as we
vary $R$.  This amounts to rescaling (\ref{curve}) by a factor of
$\bar{r}/\cosh(b/R)$:
\begin{equation}
X_1 =  {\bar{r}\cosh(X_3/R) \cos(X_{10}/R) \over \cosh(b/R)},\qquad
X_2 =  {\bar{r}\sinh(X_3/R) \sin(X_{10}/R) \over \cosh(b/R)}
\label{curve2}
\end{equation}
With this chose of scales, $X_1$ and $X_2$ goes to in the $R
\rightarrow 0$ limit for range of $X_3$ between $-b$ and $b$.  This is
the string stretching between two D2-branes.  For $X_3$ slightly
greater than $b$, both $X_1$ and $X_2$ diverges.  This indicates the
presence of D2-brane at $X_3=\pm b$.

It is also interesting to examine the asymptotic expansion of the
curve (\ref{curve2}) in the small $R$ limit.  Such an expansion
captures the information accessible from the weak coupling analysis.
The $X_2=0$ cross section of the curve (\ref{curve2}) is then given by
$$X_3 = R \cosh^{-1}( X_1 \cosh(b/R)/\bar{r})$$
which admits an asymptotic expansion
$$X_3 = b + R \ln(\bar{r}/r)+ R\left(1-{\bar{r}^2\over r^2}\right)
e^{-2b/R} + R \left( {1 \over 2} - {4 \bar{r}^2 \over r^2} +{3
\bar{r}^4 \over 2r^4 }\right) e^{-4b/R} + \ldots\, .$$
Similarly, the $X_1=0$ cross section 
$$X_3 = R \sinh^{-1}( X_1 \cosh(b/R)/\bar{r})$$
admits an asymptotic expansion
$$X_3 = b + R \ln(\bar{r}/r)+ R\left(1+{\bar{r}^2\over r^2}\right)
e^{-2b/R} + R \left( {1 \over 2} + {4 \bar{r}^2 \over r^2} +{3 \bar{r}^4 \over 2r^4 }\right) e^{-4b/R} + \ldots\, .$$
In the $R\rightarrow0$ limit, these expressions collapses to $X_3=b$
as it should. It is also interesting to note that a perturbative
analysis would have only revealed the presence of the constant term
and the logarithmic term. This is exactly the BPS configuration of the
form $y = c_p/r^{p-2}$ described in \cite{CM97,Gibbons97} for
$p=2$. At this order, the geometry is rotationally invariant in the
$(X_1,X_2)$-plane.  All the other correction appears to be suppressed
by factors of the form $e^{-b/R}$. Such terms would have only been
visible through non-perturbative analysis of the brane world volume
field theory.  It is also interesting to note that at the
non-perturbative level, rotational symmetry in $(X_1,X_2)$-plane is
broken. This is simply the manifestation of asymmetry between
$\cosh(X_3/R)$ and $\sinh(X_3/r)$ cross sections for the M-theory
membrane configuration (\ref{curve2}).

\section{Brane configurations in type IIB theory}

The method of generating intersecting brane configuration as a limit
of M-theory membrane configuration is remarkably powerful, where by
taking advantage of holomorphicity, one is able to derive the full
non-perturbative structure of the brane world volume field
configuration.  Unfortunately, this technique is limited in its
applicability to class of configurations which admits a description in
terms of either single membrane or a five-brane.  In particular, it is
not clear if this technique can be applied to study the configuration
of our interest, namely string stretched between a pair of D3-branes
in the type IIB theory.  Also, in brane configuration where a
fundamental string stretches between a pair of D-branes, the geometry in
the weak coupling limit is flat as we saw in the case of D2-branes in
the previous section. This is to be expected since the tension of
D-branes grows like $1/g$ in the weak coupling limit, whereas the
fundamental string tension is invariant under the change in the
strength of the coupling.  In type IIB theory, on the other hand, one
can keep the non-trivial geometry of branes intact even in the weak
coupling limit by stretching a D-string instead of a fundamental
string, since the tension of the D-string will also scale like $1/g$
in that limit.  In this section, we will examine this type of
configurations in type IIB theory.

Ideally, one would like to study this problem using the full
non-abelian Born-Infeld action.  Presently, we are unable to carry out
this program due to our inability to write down the non-abelian
Born-Infeld action explicitly \cite{NDBI,akiwati}.  However, for BPS
states, the solution to the Maxwellian truncation of the abelian
Born-Infeld theory is also the solution of the full Born-Infeld theory
as was shown in \cite{CM97,Gibbons97}.  It is reasonable to expect
similar relation to hold between non-abelian Born-Infeld theory and
its Yang-Mills truncation (See \ref{app1}).  Under this
assumption, we will infer the geometry of the D3-branes from the BPS
field configuration of Yang-Mills theories. Once these field
configuration for Yang-Mills theories are identified, we will be able
to compute their energies and verify the interpretation that they are
being pulled by strings of appropriate tensions.

Let us begin by considering the situation where a D-string stretches
between a pair of D3-branes, separated by a distance $2b$.  On the
D3-brane world volume, this configuration is described by a magnetic
monopole solution for $U(2)$ gauge theory coupled to adjoint Higgs
field. The eigenvalues of the adjoint Higgs field at infinity will
correspond to the distance separating the D3-brane at infinity.

Our task of determining the bent geometry of D3-branes is simplified
dramatically in light of the fact that the field configuration of a
magnetic monopole in $U(2)$ Yang-Mills theory is known explicitly in
the form of the Prasad-Sommerfeld solution \cite{PS75}. We can simply
read off the D3-brane geometry from the field configuration of the
Higgs scalar in the solution of \cite{PS75}.

The magnetic monopole of Prasad and Sommerfeld is a classical solution of 
the theory defined by an action
\begin{equation}
S = -{1 \over e^2} \int d^4x\ {1 \over 4} (F_{\mu \nu}^a)^2 + \frac{1}{2} (D^\mu \varphi^a)^2
\label{yangmills}
\end{equation}
where
$$F_{\mu\nu}^a = \partial_\mu A_\nu^a  - \partial_\nu A_\mu^a  + \epsilon^{abc} A_\mu^b A_\nu^c, \qquad D_\mu \varphi^a = \partial_\mu \varphi^a + \epsilon^{abc} A_{\mu}^b \varphi^c$$
The solution is given in terms on an ansatz
\begin{eqnarray}
A_i^a &=& \epsilon_{aij} \hat{r}_j[1-K(r)]/r \nonumber \\
A_0^a &=& \hat{r}_a J(r) / r  \label{ansatz}\\
\varphi^a & = & \hat{r}_a H(r)/r \nonumber
\end{eqnarray}
with $K(r)$, $J(r)$, and $H(r)$ for the magnetic monopole given by the
solution
\begin{eqnarray}
K(r) & = & Cr/\sinh(Cr) \nonumber \\
J(r) &=& 0  \label{monopole}\\
H(r) & = & C r \, \coth(Cr)-1 \nonumber
\end{eqnarray}
where $C$ is a constant. In order to interpret this field
configuration as a solution to the D-brane world volume theory, we
need a precise dictionary of field variables. In order to find such a
dictionary, consider expanding the abelian Born-Infeld action for the
D3-brane
\begin{eqnarray}
S &=& T_3 \sqrt{-\det(\eta_{\mu \nu} + T^{-1} F_{\mu \nu} + \partial_\mu X \partial_\nu X)} \approx T_3 \left[{1\over 4} T^{-2} F^2 +{1 \over 2} (\partial X)^2\right] \nonumber \\
& =& T_3 T^{-2} \left( {1 \over 4} F^2 + {1 \over 2} T^2 (\partial X)^2\right) \label{bi}
\end{eqnarray}
Therefore, the appropriate identification is to take
$$\varphi(r) = T X(r), \qquad {1 \over e^2} = T_3 T^{-2}$$
The adjoint Higgs scalar on the D3-brane world volume
therefore takes the form
$$X(r) = T^{-1} \tau^a  \hat{r}^a H(r)/r  = \tau^a \hat{r}^a T^{-1} \left(C\, \coth(Cr) - {1 \over r}\right)$$
where $\tau^a$ are the generators for the $SU(2)$ Lie algebra.

The form of $X(r)$ found above contains components in all directions
in the space of $SU(2)$ Lie algebra.  One would like to diagonalize
this field so that the eigenvalues can be interpreted as the position
of the D3-brane. Such a diagonalization can be carried out on an
arbitrary coordinate patch by performing appropriate gauge
transformation\footnote{In attempting to perform such a gauge
transformation globally, one encounters a Dirac-string singularity
characteristic of magnetic charges. This can be thought of
heuristically as the continuation of the D-string.}. At least locally,
we find that the embedding of D3-brane world volume into transverse
space is specified according to
$$X(r) = \pm  T^{-1} \left(C\, \coth(Cr) - {1 \over r}\right).$$
As $r$ is taken to infinity, $X(r)$ approaches $T^{-1} C$.  We can
therefore set the value of $C$ in terms of distance $\Delta X = 2b$
between the D3-branes as $C = T \Delta X/2$. The leading asymptotic
dependence of $X(r)$ on $r$ goes as $T^{-1}/r$. Except for the factor
of $g$ which accounts for the difference between the fundamental
string and a D-string, this is precisely equivalent to the field
configuration $X(r)=c_p/r^{p-2}$ encountered earlier in the context of
BPS configurations in abelian Born-Infeld theories.

\begin{figure}
\centerline{\psfig{file=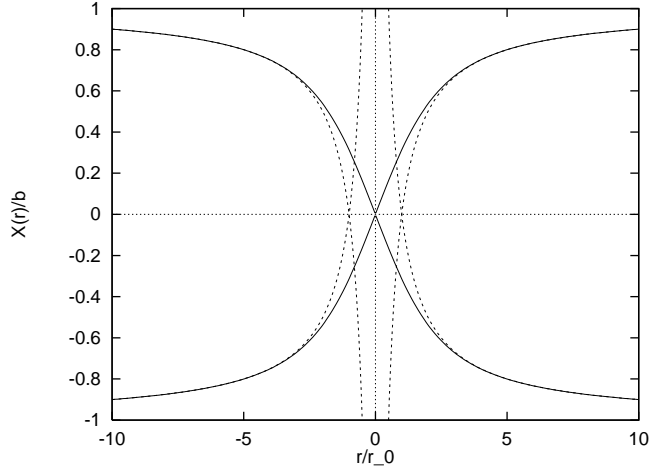,width=3.5in}}
\caption{Deformation of D3-brane by the tension of a string stretching
between them. The solid line is the shape inferred from the field
configuration of the Higgs scalar in the Prasad Sommerfeld
solution. The dotted line is the BPS configuration $X = c_p/r^{p-2}$
for the abelian Born-Infeld theory. The scale factor $r_0$ is $g/Tb$
for the fundamental string and $1/Tb$ for the D-string.
\label{fig5} }
\end{figure}

Near $r=0$, $X(r)$ behaves like
$$X(r) \approx \frac{1}{12}T (\Delta X)^2 r$$
The expectation therefore is for radius of the D3-brane spike to
shrink linearly to zero as one approaches the point halfway between the
two branes. The form of $X(r)$ is illustrated in figure \ref{fig5}.

Given the exact field configuration, one can immediately calculate its
energy.  The canonical Hamiltonian energy of  (\ref{yangmills}) for the ansatz (\ref{ansatz}) is given by
$$E = {1 \over e^2}\int d^3 r \ \left( K'^2 + {(K^2-1)^2 \over 2r^2} - {J^2 K^2 \over r^2}-{(r J'-J)^2 \over 2r^2} + {H^2K^2 \over r^2} + { (r H'-H)^2 \over 2r^2} \right)$$
For the monopole solution (\ref{monopole}), this is simply
$$E = {4 \pi C \over e^2} =4 \pi (T_3 T^{-2}) (T \Delta X /2) = {T \over
g} \Delta X.$$
This is precisely the energy of D-string stretching for the length
$\Delta X$!  The fact that this energy worked out to an expected value
suggests that we have found the correct configuration of D3-branes
sustaining precisely the amount of stress needed to balance the
tension of the D-string attached to its world volume.

Let us now discuss the situation where a fundamental string stretches
between a pair of D3-branes. The most efficient method for studying
this situation is to act on (\ref{bi}) by electric-magnetic duality
transformation. Then, after appropriate field redefinitions, the same
solution (\ref{monopole}) can be used to describe the configuration of
interest.

Under electric-magnetic duality, the action (\ref{bi}) transforms
according to rules \cite{Tseytlin96}\\
\centerline{\parbox{\hsize}{
\begin{eqnarray*}
T_3 = \frac{1}{2 \pi g} T^2 & \rightarrow & \frac{1}{2 \pi} T^2 g \\
\sqrt{-\det(\eta_{\mu \nu} + \partial_\mu X \partial_\nu X + T^{-1} F_{\mu \nu})} & \rightarrow & \sqrt{-\det(g^{-1} \eta_{\mu \nu} + g^{-1} \partial_\mu X \partial_\nu X + T^{-1} F_{\mu \nu})}\\
&& \approx T^{-2} \left( \frac{1}{4} F^2 + \frac{1}{2} T^{2} g^{-2} (\partial X)^2 \right).
\end{eqnarray*}}}
So the dual action  becomes
$$S = {1 \over 2 \pi} g \left(\frac{1}{4} F^2 + \frac{1}{2} T^{2} g^{-2} (\partial X)^2 \right)$$
with the identification 
$$\varphi(r) = T g^{-1} X(r), \qquad C = T g^{-1}\, \Delta X /2,
\qquad {1 \over e^2} = {1 \over 2 \pi} g.$$
This time, the energy is 
$$E =   {4\pi C \over e^2}  = T \Delta X$$
which is precisely the energy of fundamental string stretching for a
length $\Delta X$.

It is interesting to compare the shape of the deformed D3-brane due to D-string and the fundamental string. For the D-string we found 
\begin{equation}
X_D(r) = b\, \coth(T b r) - T^{-1}\frac{1}{r}
\label{Dconfig}
\end{equation}
whereas for the fundamental string, we found
\begin{equation}
X_F(r) = b\, \coth(T b r/g) - g T^{-1}\frac{1}{r}
\label{Fconfig}
\end{equation}
where $b$ is the position of D3-brane at infinity. The main difference
between the two field configurations is its dependence on the
coupling.  For the D1-string, the shape of D3-brane is independent of
$g$. Therefore, the bending will persist in even in the $g \rightarrow
0$ limit. For the fundamental string, on the other hand, $X_F(r)
\rightarrow b$ as $g \rightarrow 0$. the asymptotic expansion near
$g=0$ is
$$X_F(r) = b + g T^{-1}\frac{1}{r} + 2 b\, e^{-2Tbr/g} +2 b \, e^{4Tbr/g} + \ldots\ .$$
Just as in the case of D2-brane in type IIA theory, we find that only
the first two terms in this series is perturbative, and to this level
of approximation, the field configuration is identical to the form
$c_p/r^{p-2}$ discussed earlier. The problem that such a configuration
was unsuitable for describing strings stretching between pair of
D3-brane appears to be resolved entirely through non-perturbative
effects.

Finally, let us comment on the issue of whether brane configuration
(\ref{Dconfig}) and (\ref{Fconfig}) is physically distinguishable from
the configuration $c_p/r^{p-2}$ of \cite{CM97,Gibbons97} in light of
the fact that our ability to probe short distances is limited by the
length-scales of the probe itself.  To facilitate the comparison, we
also included the plot of $c_p/r^{p-2}$ in figure \ref{fig5}.  At the
point where the two branches meet, configurations (\ref{Dconfig}) and
(\ref{Fconfig}) differs from $c_p/r^{p-2}$ by a quantity of order
$r_0$.  Since $r_0$ is different for the fundamental string and the
D-string, let us examine each cases separately.

For the case of fundamental string stretching between the D3-branes,
$r_0=g/Tb$.  In the weak coupling limit, this is smaller than the
characteristic thickness of the fundamental string itself.  The
difference between (\ref{Fconfig}) and $c_p/r^{p-2}$ is buried within
the stringy/brany ``fuzz'' and appears not to have observable
consequences.

In the case of D-string stretching between the D3-branes, on the other
hand, $r_0=1/Tb$ is independent of $g$. Although $r_0$ is of order
string scale and appears to hide beneath the stringy halo, it is much
larger in comparison to the length scales accessible by say a D-string
probe which goes as $\sqrt{g/T}$ \cite{DKPS96} in the weak coupling
limit.  We therefore conclude that the feature of (\ref{Dconfig}) near
$r=0$ is real and observable, at least until the radius $r$ shrinks to
the characteristic thickness of the D-string $\sqrt{g/T}$.

\section{Discussion}

In this article, we examined the effect of bending of branes when a
string attaches one of its endpoints and pulls by its own tension.
For a system consisting of a single D-brane being pulled by a string
which extends semi-infinitely, one can analyze its static equilibrium
using abelian Born-Infeld action, as was done in
\cite{CM97,Gibbons97}. The picture which emerges from such an analysis
is that the string tension is strong enough to pull the brane world
volume all the way to infinity, provided the string coupling constant
is non-zero.

The picture needs to be modified somewhat in order to discuss the
geometry of a string stretching between a pair of D-branes, causing
the branes to bend.  In order to analyze this system along the lines
of \cite{CM97,Gibbons97}, full non-abelian generalization of the
Born-Infeld action is required.  Currently, we are unable to carry out
such a program due to our inability to write down such an action (See
\ref{app1}). In this article, we side-stepped this problem by 1)
considering a system of parallel D2 brane supporting a stretched
string which admits a full non-perturbative treatment in the M-theory
limit, and 2) considering the Yang-Mills truncation of the ``would
be'' Born-Infeld action and assuming that for BPS configurations,
Yang-Mills theory will give the correct answer. Here, we took
advantage of the known explicit forms of SU(2) magnetic monopole field
configuration. In particular, we were able to read off the embeddings
of the D3-brane world volume into spacetime directly from the field
configuration of the adjoint scalar field of the magnetic
monopole. Both of these approaches have allowed us to construct
explicit expressions describing the shape of the D2-branes and
D3-branes in type IIA and IIB theories, respectively.  In the case of
a fundamental string stretching between the branes, however, the
effect of bending becomes arbitrary small in the weak coupling limit
and the geometry is essentially trivial. For the type IIB theory,
however, one can suspend a D-string instead of a fundamental
string. The non-trivial bending persists to the weak coupling limit
and is physically observable from scattering of brane probes.

The approach of inferring brane configurations form the field
configurations of magnetic monopole solutions can be generalized to dyon
solutions, multi-monopole solutions, and larger gauge groups. These
should correspond to $(p,q)$ strings, multiple strings, and multiple
D3-branes, respectively, in the D-brane language.  One could also
imagine stretching fundamental strings for D-branes other than the
D3-branes. However, in the weak coupling limit, all non-trivial
geometry is expected to be hidden inside a stringy halo.

Although we presented the M-theory technique for describing the type
IIA D2-branes and the magnetic monopole technique for describing the
type IIB D3-branes as independent methods, the two are clearly
related, since one can map from one to the other by chain of
dualities.  These systems are also closely related to pair of
orthogonal D2-branes intersecting at a point whose BPS deformation
also admits simple holomorphic descriptions \cite{CM97}. The main
difficulty of applying the language of holomorphic curves to D3-branes
is the fact that such curves are naturally even dimensional and one
must perform at least one T-duality transformation which obscures the
distinction between spatial and gauge configurations. Yet there is a
sense in which algebraic curves appears in the construction of
magnetic monopoles \cite{Hitchin83,Hurtubise89,DonaldKron}. It would be very
interesting to understand the unifying structure which underlies these
different approaches.

\section*{Acknowledgments}

I would like to thank Per Berglund, John Brodie, Sean Carroll, Steve
Giddings, David Gross, Hiroshi Ooguri, Gary Horowitz, Juan Maldacena,
Rob Myers, Amanda Peet, Phillipe Pouliot, and Washington Taylor for
comments and useful discussions. This work was supported in part by
the NSF grant PHY94-07194.

\appendix

\section{Monopole solutions of non-abelian Born-Infeld action \label{app1}}

Throughout this article, we referred to the non-abelian generalization
of the Born-Infeld action as the necessary framework for studying the
geometry of parallel branes suspending a string.  In generalizing the
Born-Infeld action to its non-abelian counterpart, one must provide
additional information specifying the order in the product of
non-commutative factors. Attempts have been made by many authors to
resolve this ambiguity \cite{fradtseyt85,tseytlin86,argnap89}, but
only terms up to order $F^4$ were properly understood.

Recently, Tseytlin has proposed an explicit form for the non-abelian
Born-Infeld action \cite{NDBI}
\begin{equation}
S = T_p \int d^{p+1}x \, {\rm STr} \sqrt{-\det (\eta_{\mu \nu} +
T^{-1}F_{\mu \nu})} \label{ndbi}
\end{equation}
where the determinant is over the Lorenz indices $(\mu,\nu)$ and
``STr'' indicates that the trace over gauge indices is to be taken
after symmetrizing over all permutations of the non-commutative
products. Although this action has passed a number of consistency
checks, a discrepancy was noted in \cite{akiwati} for the fluctuation
spectrum of this model around a background of constant magnetic flux.
Some modification of (\ref{ndbi}) appears to be necessary in order to
resolve this discrepancy in full.

Although it is not clear how (\ref{ndbi}) will be modified when the
discrepancy of \cite{akiwati} is fully resolved, this action appears
to already exhibit many desirable features.  For example, abelian
Born-Infeld action has the property that a BPS configuration of its
Maxwellian truncation satisfies the equation of motion of the full
Born-Infeld action.  This led us to conjecture that a BPS solution of
Yang-Mills action is also a BPS configuration of the non-abelian
Born-Infeld action. It turns out that this property holds for the
non-abelian Born-Infeld action of Tseytlin (\ref{ndbi}), as we will
demonstrate in this appendix.\footnote{I am especially indebted to Rob
Myers for extended discussions on the contents of this appendix.}

The world volume theory on D3-brane contains 6 adjoint scalars.  For
our purpose, we can set all but one of these adjoint scalar to zero.
A convenient way to write down the DBI action then is to think of it as a dimensional reduction of 4+1 dimensional Born-Infeld action
$$
S={\rm STr }\sqrt{-\det(\eta_{ab} + T^{-1} F_{ab})}
\label{action}
$$
where $(a,b)$ goes from 0 to 4 and
$$F_{ab} = \left(
\begin{array}{ccccc}
0 & F_{01}& F_{02} & F_{03} & T D_0 X\\
-F_{01} & 0 & F_{12} & F_{13} & T D_1 X \\
-F_{02} & -F_{12} & 0 & F_{23} & T D_2 X \\
-F_{03} & -F_{13} & -F_{23} & 0 & T D_3 X \\
-TD_0 X  & -T D_1 X & - T D_2 X & - T D_3 X & 0 
\end{array}
\right).$$
For the static magnetic monopole solution, we can set $F_{0i}$ and $T
D_0X$ to zero, and the action reduces to the determinant of lower
right $4 \times 4$ block. Now, the magnetic monopole solution of
Prasad and Sommerfeld satisfies the identity
$$B_i = \pm {1 \over 2}\epsilon^{ijk} F_{jk} = \pm T D_i X$$
or equivalently
$$
F_{ij} = \pm {1 \over 2} \epsilon_{ijkl}F^{kl}.
$$
This is the familiar BPS condition which follows from supersymmetry.

In deriving the equation of motion for (\ref{ndbi}), we follow the
approach of \cite{akiwati} where we take the expression inside the
symmetrized trace as if they are abelian until the end of the
calculation, at which point we will assign a singe power of $\tau^a$
for each factor of $F$ and sum over traces of all permutations of the
non-abelian factors. The variation of  (\ref{ndbi}) with respect to the 
fields inside the symmetrized trace is simply
\begin{equation}
\delta S ={\rm STr} \left[ \sqrt{-\det(\eta_{ab}+T^{-1} F^0_{ab})}  (\eta+T^{-1}F^0)^{-1}_{ab} (D^a \delta A^b - D^b \delta A^a) \right].
\label{variation}
\end{equation}
Self-dual field configuration has a unique property that the
determinant factor is a perfect square:
$$-\det( \eta_{ab} + T^{-1} F^0_{ab}) = \left(1 + (T^{-1} F^0_{12})^2 + (T^{-1} F^0_{13})^2 + (T^{-1} F^0_{23})^2\right)^2.$$
Combining this with the fact that 
$$
{1\over 2} \left( (\eta + T^{-1} F^0)^{-1}_{ab} - (a \leftrightarrow
b) \right) = {T^{-1}F_{ab} \over 1 + (T^{-1} F^0_{12})^2 + (T^{-1}
F^0_{13})^2 + (T^{-1} F^0_{23})^2}$$
allows us to simplify (\ref{variation}) to read
$$\delta S  = {\rm STr} \left[ T^{-2} F^0_{ab} D^a \delta A^b \right].
$$
At this point, since we only have two factors of the non-commutative
elements, there is no need to distinguish the symmetrized trace and
the ordinary trace.  Now one can apply ordinary integration by parts
to obtain the equation of motion
$$ D^a F^0_{ab} = 0,$$
which is precisely identical to the equation of motion for the
Yang-Mills action.  Since the solution of Prasad and Sommerfield
satisfies the equation of motion of the Yang-Mills action exactly, and
since it is a self-dual field configuration, it follows that it also
satisfies the equation of motion of (\ref{ndbi}) exactly.

Although we focused on purely magnetic solutions, one can generalize
this argument to the dyon solution where the electric field is also
turned on \cite{PS75}. Such a solution is best thought of as the boost
of the magnetic monopole solution in the 04-plane. It is trivial to
repeat the argument replacing $F_{ab}$ with $(\Lambda^{-1} F
\Lambda)_{ab}$.

It would be extremely interesting to understand the resolution of
discrepancy noted in \cite{akiwati}, and how such a resolution affects
the result of this appendix.

\bibliographystyle{plain} \bibliography{main}

\end{document}